\documentclass[journal=jctcce,manuscript=article]{achemso}

\usepackage{amsmath}
\usepackage{amssymb}
\usepackage{graphicx}

\usepackage{dcolumn}
\usepackage{bm}
\usepackage{longtable}
\usepackage{setspace}
\usepackage[usenames, dvipsnames]{color}
\usepackage{subfigure}
\usepackage{multirow}
\usepackage{dsfont}
\usepackage[normalem]{ulem}
\usepackage{lipsum}
\usepackage{xcolor}

\DeclareMathOperator*{\argmin}{argmin}

\newcommand{\red}[1]{\textcolor{black}{#1}}

\author{Ivan Duchemin}
\email{ivan.duchemin@cea.fr}
\affiliation{Univ. Grenoble Alpes, CEA, IRIG-MEM-L\_Sim, 38054 Grenoble, France}
\author{Xavier Blase}
\affiliation{Univ. Grenoble Alpes, CNRS, Inst NEEL, F-38042 Grenoble, France}

\title{  Cubic-scaling all-electron $GW$ calculations with a separable density-fitting space-time approach }

\date{\today}
\begin{document}


\begin{abstract}
We present an implementation of the $GW$ space-time approach  that allows cubic-scaling all-electron calculations with standard Gaussian basis sets without exploiting any localization nor sparsity considerations. The independent-electron susceptibility is constructed in a time representation over a non-uniform distribution of real-space locations $\lbrace {\bf r}_k \rbrace$ optimized     within a separable resolution-of-the-identity framework to reproduce   standard Coulomb-fitting calculations with meV accuracy.    The compactness of the obtained $\lbrace {\bf r}_k \rbrace$ distribution leads to a crossover with the standard Coulomb-fitting scheme for system sizes  below a few hundred  electrons.  The needed analytic continuation follows  a recent approach that requires the continuation of the screened Coulomb potential rather than the much more structured self-energy. The present scheme is benchmarked over large molecular sets and scaling properties are demonstrated on a family of  defected hexagonal boron-nitride flakes containing up to 6000 electrons. 
\end{abstract}



 \section{Introduction}

The $GW$ approximation \cite{Hed65,Str80,Hyb86,God88,Farid88,Ary98,Farid99,Oni02,Pin13,ReiningBook,Gol19rev} to the exchange-correlation self-energy has become a standard approach in solid-state physics to explore the electronic properties  of metallic or semiconducting materials. Its accuracy was indeed proven superior to standard DFT calculations relying on the Kohn-Sham \textit{ansatz} for the electronic energy levels (for large benchmark calculations on inorganic crystals, see e.g. Refs.~\citenum{Schilf06,Shishkin07}). Further, following early applications in the late 90s, \cite{Ethridge_1996,Horst_1999,Rohlfing_1999} the $GW$ formalism is  nowadays widely used as well  for the study of gas phase or dense ordered or disordered organic molecular systems. \cite{Stan_2006,Sai_2008,Ma_2009,Rostgaard10,Blase_2011,Faber_2011,Faber_2011b,Foerster_2011,Ke2011,Baumeier_2012,Korzdorfer12,Bruneval13,Pham_2013,Setten_2013,Umari_13,Cudazzo_13,Lischner14,Koval14,Krause15,Setten_2015,Kaplan16,Wilhelm16,Rangel_2016,Scher16,Knight_2016,Vlcek17,Maggio17,Maggio17b,Marom_2017,Golze18,Govoni18,Veril18,Wehner18,Holzer19,Bruneval19,Li_2019,Koval19,Bruneval20,Loos20,Berger21} The development of codes exploiting standard   Gaussian  atomic basis sets   allowed in particular  the comparison of all-electron  $GW$ calculations with higher level quantum-chemistry techniques (e.g. coupled-cluster) performed with the very same running parameters (geometry, atomic basis sets, resolution-of-the-identity, etc.) \cite{Bruneval13,Krause15,Rangel_2016,Knight_2016}

The scaling of the number of operations needed to perform $GW$ calculations with respect to the system size is typically   $\mathcal{O}(N^4)$ within traditional planewave implementations. This scaling can be preserved with localized basis sets provided that resolution-of-the-identity (RI) techniques 
\cite{Whitten_1973,Baerends_1973,Dunlap_1979,Vahtras_1993,Klopper_2002,Ren_2012,Duchemin_2017} are used to avoid calculating response functions, such as the susceptibility, in the product space associated with valence-to-virtual molecular orbital products. While such a moderate scaling already allows calculations on systems containing well over a hundred atoms on supercomputers, \cite{Duchemin_2012,Govoni_2015,Li_2017,Ben_2019} attempts to deliver $GW$ calculations with a lower scaling appeared with the seminal space-time approach by Rojas, Needs, Godby in  1995  \cite{Roj95} and are now blooming. \cite{Foerster_2011,Neu13,Liu16,Vlcek17,Wil18,Gao20,Kim20,Kup20,Forster20,Wilhelm21}

This space-time formalism \cite{Roj95} stands as the first cubic-scaling $GW$ approach relying on the separability of the independent-electron susceptibility $\chi_0$ as the product of two Green's functions when expressed over a real-space grid, adopting further a time representation.  This factorisation allows   decoupling the summation over occupied and virtual molecular orbital contributions, leading to a cubic scaling scheme instead of the traditional quartic scaling calculation of $\chi_0$. Such a reduced scaling does not rely on any localization nor sparsity considerations associated with e.g. 3-center integrals within the local direct overlap metric \cite{Foerster_2011,Wil18,Forster20} or a range-truncated  Coulomb metric.  \cite{Wilhelm21} 

The imaginary-time formulation at the heart of the $GW$ space-time approach  is identical to the Laplace transform idea  already in use in quantum chemistry for e.g. MP2 calculations. 
\cite{Almlof91,Haser92} On the contrary, the use of a real space-grid  was more naturally rooted in the pseudopotential planewave   community, allowing by Fourier transform of the planewave basis to obtain relatively sparse uniform real-space  grids. The space-time $GW$ approach was more recently adapted to a full potential projector-augmented wave methodology,  \cite{Liu16} building on an earlier application to calculating RPA correlation energies with cubic scaling. \cite{Kaltak14}

In the case of all-electron calculations, the size of the real-space grid may seem an \textit{a priori}  bottleneck. However, real-space quadrature strategies have been already developed with much success in quantum chemistry for accelerating the calculation of 2-electron Coulomb integrals, including the  chain-of-sphere (COSX) semi-numerical approach to exchange integrals \cite{Neese09,Neese11,Klopper13} or the general tensor hypercontraction  mathematical framework in its specific  least-square grid optimization implementation (LS-THC). \cite{Parrish12,Todd12,Todd2015} More recently, the  interpolative separable density fitting (ISDF) approach \cite{Lu15,LU16} represents a versatile strategy to combine the standard quantum chemistry resolution-of-the-identity (RI) techniques 
with a separable  representation of the coefficients of molecular orbital products over auxiliary basis sets. The ISDF approach is now   developing   in the pseudopotential planewave or real-space grid community, \cite{Hu17,Dong18,Hu18,Hu20,Gao20} including a recent $GW$ implementation. \cite{Gao20}
Similarly, building on the expertise with resolution-of-the-identity (RI) techniques and/or real-space quadratures for Coulomb integrals, the ISDF scheme is also being explored by the quantum chemistry community working with localized (e.g. Gaussian) basis sets for explicitly correlated all-electron calculations such as QMC or M\"{o}ller-Plesset   techniques. \cite{Malone19,Lee20}


In a recent study, we presented an alternative to the ISDF formalism applied to all-electron Hartree-Fock, MP2 and RPA calculations with  Gaussian basis sets. \cite{Duc19} In this scheme, standard  auxiliary $\lbrace P_\mu \rbrace$ basis sets (e.g. cc-pVXZ-RI \cite{Weigend02} or def2-XZVP-RI \cite{Weigend98}) are provided as an input, as in any standard resolution-of-the-identity (RI) calculation, but the fitting procedure takes as an intermediate the expression of wave function and densities over compact non-uniform real-space grids $\lbrace {\bf r}_k \rbrace$.  The corresponding fitting weights result  from solving a quadrature equation that aims at reproducing the results of a standard Coulomb-fitting (RI-V) calculation. Adopting the space-time approach,  cubic-scaling calculations of the independent-electron susceptibility at imaginary frequencies, and resulting RPA correlation energy, could be achieved.
As compared to the corresponding RI-V calculation, an accuracy of a few $\mu$Hartree/electron  was demonstrated for RPA correlation energies with  $\lbrace {\bf r}_k \rbrace$ distributions typically   4 times larger than the input auxiliary basis, allowing a crossover with the standard quartic-scaling RI-V RPA calculations for systems of the size of pentacene.

In the present work, we extend this formalism to all-electron cubic-scaling $GW$ calculations.    In contrast with the RPA formalism, where only imaginary-frequency susceptibilities are needed, we further exploit a recently developed \cite{Fri19,Duc20}  analytic continuation scheme that brings to the real-frequency axis  the dynamically screened Coulomb potential $W$  rather than  the much more structured $GW$ self-energy. As compared to standard   RI-V $GW$ calculations, we demonstrate an accuracy at the meV level for the  quasiparticle energies of large molecular sets. Finally, cubic scaling is evidenced using a family of defected hexagonal boron-nitride flakes with increasing radius containing up to 6000 electrons, with a crossover with the standard quartic scaling RI-V $GW$ calculations for systems containing a very few hundred electrons. 

\section{Theory}

\subsection{ The $GW$ formalism with resolution-of-the-identity }

We start by describing the standard resolution-of-the-identity (RI) framework for $GW$ calculations. A more detailed discussion on RI techniques applied to MBPT can be found in Ref.~\citenum{Ren_2012}. We just recall here that the essence of RI approximations, developed in particular to tackle the calculation of 2-electron 4-center Coulomb integrals with localized atomic orbital (AO) basis sets, \cite{Whitten_1973,Baerends_1973,Dunlap_1979,Vahtras_1993,Klopper_2002,Ren_2012,Duchemin_2017}  amounts to expressing the product of 2 molecular orbitals (MOs) over an auxiliary basis set $\lbrace P_{\mu} \rbrace$,   namely:
\begin{equation} \label{eq:RIgeneral}
\phi_n({\bf r}) \phi_m({\bf r}) = \sum_{\mu} \mathcal{F}_{ \mu } (\phi_n \phi_m) P_{\mu}({\bf r})
\end{equation}
where we work with finite size systems allowing real molecular orbitals. 
For localized atomic-orbitals (AO) basis   calculations, the auxiliary basis is typically 2-3 times larger   than the AO basis set used to expand the MOs.  As an example, the accurate RI-V Coulomb-fitting approach\cite{Vahtras_1993} defines the coefficients $\mathcal{F}_{\mu}$ as:
\begin{equation} \label{eq:RIV}
\mathcal{F}^V_{\mu}(\phi_n \phi_m) = \sum_{\nu}
 [V^{-1}]_{\mu\nu} ( P_{\nu} | \phi_n \phi_m) 
\end{equation}
with $(P_{\nu} | \phi_n \phi_m)$ the 3-center Coulomb integrals
\begin{equation*}
    ( P_{\nu} | \phi_n \phi_m ) = \int d{\bf r} d{\bf r}' \; 
\frac{ P_{\nu}({\bf r}) \phi_n({\bf r}') \phi_m({\bf r}') }{ | {\bf r}-{\bf r}'|}
\end{equation*}
and $V_{\mu\nu}$ the Coulomb matrix elements in the auxiliary basis.
Coming now to the $GW$ formalism, we start with the expression of the independent-electron susceptibility along the imaginary-frequency axis:

\begin{equation}
     \chi_0({\bf r},{\bf r}' ; i\omega)  
    = 2 \sum_{ja}   \frac{   \phi_j^*({\bf r})  \phi_a({\bf r})        \phi_a^*({\bf r}')  \phi_j({\bf r}')   }{ i\omega - ( \varepsilon_a - \varepsilon_i) } + c.c.
\end{equation}
where (j) and (a) index occupied and virtual molecular orbitals (MOs), respectively, and where the factor (2) indicates a closed shell system. Expanding MO products over an auxiliary basis in the case of real-valued MOs leads to:  
\begin{equation}
        \chi_0({\bf r},{\bf r}' ; i\omega)
     = \sum_{\mu\nu} P_{\mu}({\bf r}) \cdot [\chi_0^{RI}(i\omega)]_{\mu\nu} \cdot P_{\nu}({\bf r}') 
\end{equation}
    with
\begin{equation} 
    [\chi_0^{RI}(i\omega)]_{\mu\nu}  = 2
    \sum_{ja}   \frac{ \mathcal{F}_{\mu}(\phi_j \phi_a) \mathcal{F}_{\nu}(\phi_j \phi_a)  }{ i\omega - ( \varepsilon_a - \varepsilon_j) }  + c.c.   
    \label{eq:xi0auxbasis}
\end{equation}
In the standard RI framework, it is the quantity $[\chi_0^{RI}(i\omega)]_{\mu\nu}$ that is calculated   in the auxiliary $\lbrace P_{\mu} \rbrace$ basis. The following steps start with the definition of the $GW$ correlation self-energy as a  convoluted integral along the real-energy axis   
\begin{equation}
\Sigma^C({\bf r},{\bf r}' ; \red{E} ) = \frac{i}{2\pi} 
\int_{-\infty}^{\infty} d\omega \; e^{i \omega 0^+} G({\bf r},{\bf r}';E+\omega)
\widetilde{W}({\bf r},{\bf r}';  \omega)
\end{equation}
with $\widetilde{W}=(W-V)$ and where $G$, $W$ and $V$ are the time-ordered 1-body Green's function, the screened and bare Coulomb potentials, respectively. In the contour-deformation approach, \cite{God88,Farid88} this expression is transformed into  an integral along the imaginary-energy axis, plus the contribution of a few residues involving the screened Coulomb potential $W$ calculated at  real energies:
\begin{align}  \label{eq:contourdeform}
\Sigma_C^{GW}({\bf r},{\bf r}' ; \red{E} ) &= \frac{-1}{2\pi} 
\int_{-\infty}^{\infty} d\omega \; G({\bf r},{\bf r}';E+i\omega)
\widetilde{W}({\bf r},{\bf r}'; i\omega) \\
 &- \sum_i \phi_i({\bf r}) \phi_i({\bf r}') \widetilde{W}({\bf r},{\bf r}'; \varepsilon_i -E ) \theta(\varepsilon_i -E)  \nonumber \\
 &+ \sum_a \phi_a({\bf r}) \phi_a({\bf r}') \widetilde{W}({\bf r},{\bf r}'; E - \varepsilon_a ) \theta(E-\varepsilon_a ) \nonumber
\end{align}
Expressing the Green's function in a quasiparticle form
\begin{equation}
   G({\bf r},{\bf r}' ; i\omega) = \sum_{n}
   \frac{ \phi_n({\bf r}) \phi_n({\bf r}')  }
   {  i \omega - \varepsilon_n + i \eta \times \text{sgn}(\varepsilon_n - \mu) }
\end{equation}
with $\eta=0^+$ and $\mu$ the chemical potential, it appears that the  expectation value of the $GW$ correlation self-energy $\langle \phi_n | \Sigma_C^{GW} | \phi_n \rangle$ operator   only requires   integrals of the kind:
\begin{equation}
\langle \phi_n \phi_m |  {W}(z)  |  \phi_n \phi_m \rangle = 
\sum_{\mu\nu} \mathcal{F}_{\mu}(\phi_m \phi_n)
\mathcal{F}_{\nu}(\phi_m \phi_n)
\langle P_{\mu} | W(z)  |  P_{\nu} \rangle
\end{equation}
with $z$ along the imaginary or real-energy axes. The needed    $\langle P_{\mu} | W(z)  |  P_{\mu} \rangle $ matrix elements of the screened Coulomb potential are obtained from a Dyson-like equation projected into   the auxiliary basis 
\begin{equation}
    \langle P_{\mu} | \ {W}(z)  |  P_{\nu} \rangle = 
     \langle P_{\mu} | V  |  P_{\nu} \rangle + \sum_{\zeta \rho} 
      \langle P_{\mu} | V  |  P_{\zeta} \rangle  \cdot
     [\chi_0^{RI}(z)]_{\zeta \rho} 
     \cdot \langle P_{\rho} | W(z)  |  P_{\nu} \rangle
     \label{eq:Dyzon}
\end{equation}
 where the random phase approximation (RPA) approximation is used.

\subsection{The space-time approach from a separable resolution-of-identity framework}


With the size of the auxiliary basis scaling linearly with system size, straightforward calculation of the $[\chi_0^{RI}(i\omega)]_{\mu\nu}$ matrix elements from equation \ref{eq:xi0auxbasis} requires $\mathcal{O}(N^4)$ steps. In other words, for each $(P_{\mu} , P_{\nu})$ pair a double summation over occupied and unoccupied MOs is required. Following Alml\"{o}f and H\"{a}ser, \cite{Almlof91,Haser92} the Laplace transform: 
\begin{equation}
  \frac{1}{   i\omega - (\varepsilon_a - \varepsilon_j) } + c.c.
    =   -2 \int_0^{+\infty} d\tau \; \cos(\omega \tau)
       e^{ - ( \varepsilon_a - \varepsilon_j  )   \tau} 
\end{equation}
where the time integral converges since $(\varepsilon_j  - \varepsilon_a) < 0,$ 
allows to disentangle occupied and virtual energy levels in the denominator, leading to an imaginary time formulation 
\begin{equation}
       [\chi_0^{RI}(i\tau)]_{\mu\nu}  = -
    2 i \sum_{ja}    \mathcal{F}_{\mu}(\phi_j \phi_a) \mathcal{F}_{\nu}(\phi_j \phi_a)    e^{   \varepsilon_j   \tau}  e^{ -  \varepsilon_a   \tau }    \label{eq.chi0bb}
\end{equation}
where the (i) factor is introduced to match the standard definition of the independent-electron susceptibility in the time domain. However, occupied and virtual MOs are still entangled in the
$\mathcal{F}_{\mu/\nu}(\phi_j \phi_a) $ weight factors. This is precisely the goal of the separable RI introduced in Ref.~\citenum{Duc19} in the context of RPA total energies, with the expansion :
\begin{equation} \label{eq/RIRS}
  \mathcal{F}_{\mu}^{RS}(\phi_j \phi_a) =
     \sum_k  M_{\mu k} \; \phi_j( {\bf r}_k )  \phi_a( {\bf r}_k ) 
\end{equation}
where the $\phi_j$ and $\phi_a$ MOs are factorized, leading to the wording separable-RI that we label RI-RS where RS stands for real-space. 
The present scheme targets a standard calculation with input molecular orbitals (MO) Gaussian basis   and  its associated auxiliary basis sets, and the   $\lbrace {\bf r}_k \rbrace$ distribution is an intermediate representation designed to reproduce the accuracy of a standard Coulomb-fitting calculation with the input basis sets. This is   described here below and in Ref.~\citenum{Duc19} for Hartre-Fock, RPA and MP2 calculations.

\begin{figure}[t]
	\includegraphics[width=12cm]{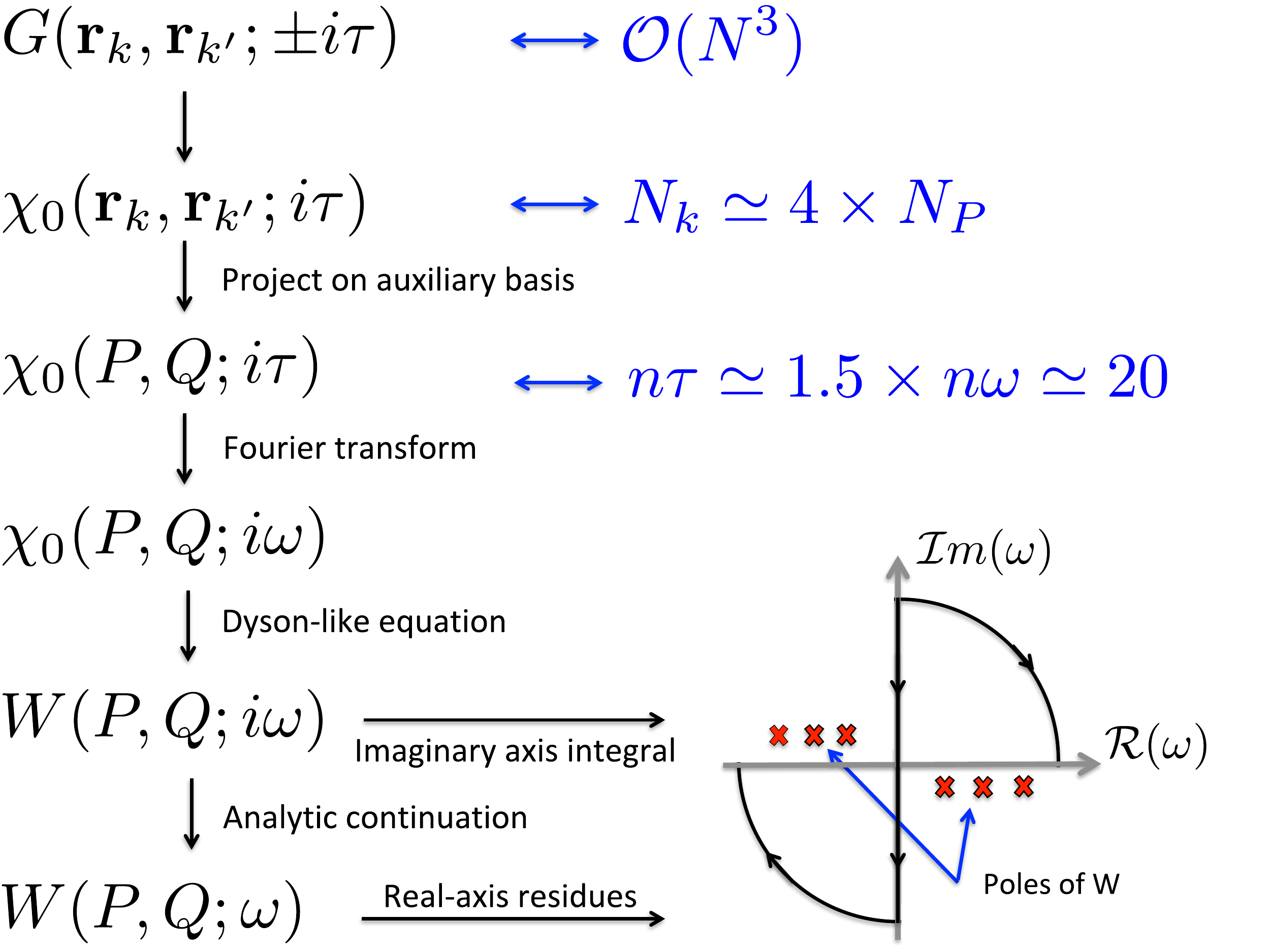}
	\caption{
		Schematic representation of the steps involved in the present cubic scaling all-electron space-time approach. The optimized set  of real-space positions $\lbrace {\bf r}_k \rbrace$  is typically  4 times as large as the corresponding input Gaussian auxiliary basis $\lbrace P_{\mu} \rbrace$. The number of imaginary times and frequencies is set by n$\tau$ and n$\omega$. In the contour deformation approach (see Inset)  only the screened Coulomb potential $W$ needs to be continued from the imaginary   to  the real-energy axis, avoiding the continuation of the much more structured self-energy (see Ref.~\citenum{Duc20}).
	\label{fig:flow}}
\end{figure}

The separable real-space RI (RI-RS) leads to  expressing the independent-electron susceptibility matrix elements in the auxiliary basis as
\begin{equation}
    [\chi_0(i\tau)]_{\mu\nu}  \stackrel{RI-RS}{=}
     \sum_{kk'}  M_{\mu k} \cdot \chi_0({\bf r}_k, {\bf r}_{k'} ; i\tau) \cdot M_{\nu k'}
     \label{auxtoaux}
\end{equation}
where for any $( {\bf r},{\bf r}' )$ pair of points in real-space
\begin{eqnarray}
\chi_0({\bf r},{\bf r}';i\tau) &=& - i G({\bf r},{\bf r}';i\tau) G({\bf r}',{\bf r};-i\tau) \label{xi0rrit} \\
G({\bf r},{\bf r}';\red{i\tau}) &=& \phantom{-}i \sum_j^{\text{occ}} \phi_j({\bf r}) \phi_j({\bf r}') e^{  \varepsilon_j \tau} \;\;\;\;  (\tau > 0)  \\
                         & =& -i  \sum_a^{\text{vir}} \phi_a({\bf r}) \phi_a({\bf r}') e^{    \varepsilon_a \tau}  \;\;\;\;  (\tau < 0)  
\end{eqnarray}
where $G$ is the time-ordered one-body Green's function. Considering here that the MOs are spin-orbitals, one can recover the factor two in Equation~\ref{eq.chi0bb} for spin-restricted systems.  Eq.~\ref{xi0rrit} is the standard equation  for the original space-time  approach\cite{Roj95}  with the summations over occupied and virtual MOs  completely decoupled, leading to a strictly cubic-scaling number of operations, independently of any localization nor sparsity properties.  After construction of the imaginary-time independent-electron susceptibility in the  $\lbrace P_{\mu} \rbrace$ Gaussian auxiliary basis following Eq.~\ref{auxtoaux}, the $[\chi_0^{RI}(i\omega)]_{\mu\nu}$ are obtained by Fourier transform at imaginary frequencies. Following Eq.~\ref{eq:Dyzon},  the  screened Coulomb potential   $\langle P_{\mu} | \ {W}(z)  |  P_{\nu} \rangle $ is finally obtained along the imaginary frequency axis.  Details about the time and frequency grids will be given below when discussing the analytic continuation of $W$ to the real-axis. The overall flow of calculations is presented in Fig.~\ref{fig:flow}.
\red{ While we use in the present scheme the analytic continuation of $W$, the standard analytic continuation of $\Sigma$ to the real axis can also be used once the $\chi_0(P,Q;i\omega)$ susceptibilities are obtained using the real-space imaginary-time approach.}

\subsection{ Construction of the $\lbrace {\bf r}_k \rbrace$ distributions } 

A crucial aspect of the present scheme is the size of the $\lbrace {\bf r}_k \rbrace$-set that  controls the prefactor associated with the present cubic-scaling scheme.
Following our implementation of cubic-scaling RPA calculations in an all-electron space-time approach, \cite{Duc19} the central idea is not to use a generic real-space grid (such as Becke grids \cite{Becke88} adopted to express densities in DFT codes) but to optimize for each atomic species a reduced set of $\lbrace {\bf r}_k \rbrace$ points sufficient to reach the \textit{accuracy of the standard Coulomb-fitting RI-V approximation} in conjunction with the chosen auxiliary basis set. 
Such task is performed by minimizing the difference between the
$\mathcal{F}^{RS}$ and $\mathcal{F}^{V}$ fitting  procedure   as defined in  Eqs.~\ref{eq:RIV} and~\ref{eq/RIRS}, in the Coulomb norm sense, and taking into account all MO products of a single atom of the species considered. As introduced in Ref.~\citenum{Duc19}, the $M_{\mu k}$ coefficients are fixed through a linear least square equation, and thus only the $\lbrace {\bf r}_k \rbrace$-sets are considered as optimization variables. The global minimization process thus writes   
\begin{equation}
    \argmin_{ \lbrace {\bf r}_k \rbrace } \; \sum_{\mu\alpha{\alpha}'} \Big| \big(\mathcal{F}_{\mu}^{RS}( \alpha{\alpha}' ) - \mathcal{F}_{\mu}^{V}(\alpha{\alpha}')\big) P_\mu  \Big|_{V}^2
    \label{fitting}
\end{equation}
where the $\lbrace \alpha  \rbrace$ are the Gaussian basis functions used to expand the MOs. 

For a given atom, the initial set of points are constructed as a superposition of high symmetry subsets of Lebedev grids up to order 9, associated with different sphere radii (see Supporting Information Ref.~\citenum{Duc19}). 
The optimization process starts by  minimizing the penalty function of Eq.~\ref{fitting} adjusting first the radii. This is similar to the optimization strategy adopted in the grid-based formulation of LS-THC  \cite{Todd2015} but fitting the codensities coefficients rather than the 4-center Coulomb integrals.   In a second step, all constraints are raised and every point is allowed to move independently.
This  non-linear minimization process is performed using a basin-hopping mechanism coupled to a L-BFGS (limited memory Broyden-Fletcher-Goldfarb-Shanno) algorithm. We emphasize that such a step is done once for all for a given element and the chosen   basis sets.
Experimenting with such a strategy for the   def2-TZVP / def2-TZVP-RI associated basis sets leads to 100 $\lbrace {\bf r}_k \rbrace$ points   for H and He, and 336, 436 and 536 points for elements in the second, third, and fourth row of the periodic table, respectively. Such grid sizes allow an agreement at the meV level between subsequent quasiparticle energies calculated with the present real-space approach and the standard Coulomb-fitting RI-V scheme. Except for the first row, this is typically 3.5 to 4.5 times larger than the number of elements in the def2-TZVP-RI set. Better optimization schemes and reducing the initial number of Lebedev subsets may lead to reduce these $\lbrace {\bf r}_k \rbrace$ distribution  sizes. However, as shown below, the present approach already provides an excellent accuracy-to-cost ratio and appears to be very robust, with no outliers as tested over large molecular sets.

In a second step, the $\lbrace {\bf r}_k \rbrace$ distribution for the molecular system is built as the superposition of the isolated atoms $\lbrace {\bf r}_k \rbrace$ distributions and only the weights $\lbrace   M_{\mu k} \rbrace$, as defined in Eq.~\ref{eq/RIRS}, need to be calculated for each considered molecular system. Such a step only requires $\mathcal{O}(N^3)$ operations since the least-square estimator matrix $[M]_{\mu k}$ is obtained in a matrix multiplication/inversion formulation from the target $\mathcal{F}^V_{\mu}$ coefficients
(see  Ref.~\citenum{Duc19}). \red{As such, the weights $\lbrace   M_{\mu k} \rbrace$ are univocally defined once the $\{ {\bf r}_k \}$ grid and $\mathcal{F}^V_{\mu}$ factors are set-up. We provide in the Supporting Information a graph confirming the cubic-scaling of the $\lbrace   M_{\mu k} \rbrace$ construction that amounts to about 25$\%$ of the total CPU time, including the calculation of the target  $\mathcal{F}^V_{\mu}$, for non-self-consistent $G_0W_0$ calculations. } This 2-step process, namely the optimization of the $\{ {\bf r}_k \}$-distribution on isolated atoms,  dramatically simplifies the minimization process while preserving excellent accuracy as demonstrated below.     

Concerning previous RPA calculations using the present separable RI with Laplace transform scheme,  \cite{Duc19} the crucial observation   was that indeed  $\lbrace {\bf r}_k \rbrace$-sets typically  4 times larger than the used auxiliary   $\lbrace P_{\mu} \rbrace$ basis set were sufficient to reproduce the accuracy of standard RI-V calculations to within a few  $\mu$Hartree/electron for the exchange, RPA and MP2 total energies. Namely, replacing the standard RI-V $\mathcal{F}_{\mu}^{V}$ coefficients by their $\mathcal{F}_{\mu}^{RS}$  approximants preserved an excellent accuracy, with a number of points sufficiently small to offer a crossover with the standard quartic-scaling RI-V RPA approach for systems of the size of pentacene. 

We will perform here below the corresponding accuracy check for the quantity of interest here, namely the $GW$ quasiparticle energies, showing that  meV accuracy can be achieved with a crossover between the present separable  RI-RS scheme and standard RI-V calculations for systems containing less than a few hundred electrons. This crossover is independent of the compactness and dimensionality of the studied systems since   sparsity and localization are not exploited.

\subsection{ Analytic continuation, frequency and time grids }

An important aspect of the space-time approach is the required analytic continuation from the imaginary to the real-frequency axis. With the calculation of the susceptibility $\chi_0(i\tau)$ at imaginary times, the imaginary-frequency $\chi_0(i\omega)$ analog can be obtained by Fourier transform. From such quantities, the screened Coulomb potential $W(i\omega)$ and self-energy $\Sigma(i\omega)$ can be obtained at imaginary frequencies \red{ and efficiently   continued analytically to the real energy axis as performed in many codes.}  

 An alternative to the analytic continuation of the self-energy was proposed by Christoph Friedrich in the context of $GT$ calculations on solid iron, \cite{Fri19} and by ourselves in the present case of $GW$ calculations on molecular systems with extensive benchmark accuracy checks.\cite{Duc20}  The central idea is to adopt the contour deformation scheme  where the quantity needed along  the real-axis is no longer the self-energy directly, but the screened Coulomb potential (see second and third lines of Eq.~\ref{eq:contourdeform}). Since the self-energy contains $ (N_W \times N_G )$ poles, where $N_G$ and $N_W$ are respectively the number of poles of the Green's function and screened Coulomb potential, the screened Coulomb potential is much less structured than the self-energy itself, leading to a much more robust   analytic-continuation scheme. Difficult test cases drawn from the $GW$100 test sets, \cite{Setten_2015} such as the $MgO$ or $BN$ dimers, were shown to be very accurately treated with the calculations of the screened Coulomb potential $W(i\omega)$ for no more than 12 frequencies along the imaginary-axis. In particular, theses frequencies are constructed so as to minimize the error over the imaginary axis integration contribution to Equation~\ref{eq:contourdeform}.
We address  the reader to Ref.~\citenum{Duc20} for a detailed presentation of this scheme. We keep the number of imaginary frequencies to $n\omega$=12 that was shown in this former study to lead to sub-meV accuracy, as compared to the contour deformation scheme, for $GW$ calculations on frontier orbitals using this ``robust" analytic continuation scheme. 
\begin{table}
\begin{tabular}{c|c|c|c|c}
n$\omega$  &     RI-V & RI-RS & RI-RS + LT \\  
\hline 
\multicolumn{4}{c}{HOMO} \\ 
\hline
6  & -7.55777 & -7.55777 & -7.55787 \\
8  & -7.56073 & -7.56072 & -7.56071 \\
10 & -7.56113 & -7.56112 & -7.56112 \\
12 & -7.56112 & -7.56111 & -7.56111 \\
14 & -7.56112 & -7.56111 & - - \\
\hline 
\multicolumn{4}{c}{LUMO} \\ 
\hline
6  & -0.75723 & -0.75719 & -0.75713 \\
8  & -0.76020 & -0.76016 & -0.76018 \\
10 & -0.76046 & -0.76042 & -0.76042 \\
12 & -0.76044 & -0.76040 & -0.76040 \\
14 & -0.76044 & -0.76040 & - - \\
  \end{tabular} 
  \caption{ Acridine def2-TZVP $G_0W_0$@PBE0 HOMO and LUMO (in eV) convergence against the imaginary frequency grid size. We keep the ratio $n\tau = 1.5 \times n\omega$.  The real-space with Laplace transform (RI-RS + LT) depends on both frequency and time grids. All calculations are performed with the auxiliary def2-TZVP-RI basis set, while RI-RS calculations use an extra $\lbrace {\bf r}_k \rbrace$ distribution optimized for the corresponding def2-TZVP/def2-TZVP-RI basis sets association. }
  \label{Table:wtgrids} 
  \end{table}
  
Once the imaginary frequencies are set, the corresponding imaginary-time grid is constructed following Ref.~\citenum{Duc19}, where the present space-time approach was explored for RPA total energy calculations. The selected times $\lbrace \tau_p, p=1,n\tau \rbrace$ are optimized for the chosen set of  imaginary frequencies   $z_k$ (k=1, n$\omega$) through the minimization process:
\begin{equation}
    \arg\min_{\omega_k^p, \tau_p} \left[ \sum_k \int^{\ln(E_{max})}_{\ln(E_{min})}   du
    \left| \sum_p \omega_k^p e^{ - \tau_p e^{u} } -
    \left[ \frac{1}{e^{u} + iz_k} - \frac{1}{e^{u} - iz_k}\right] \right|^2  \right]
\end{equation}
where $\omega_k^p$ is the weight associated with a given $\tau_p$ time   for a targeted $z_k$ frequency. The energies $E_{min}$ and $E_{max}$ are the energy gap and the maximum ($\varepsilon_a - \varepsilon_i)$ value, respectively. The $ 1/(e^{u} \pm iz_k) $ factors represent the pole structure of the independent-electron susceptibility along the imaginary axis. The $\sum_p \omega_k^p e^{-\tau_p e^{u}} $ approximant translates the fact that 
$e^{- a |\tau|}$ ($a > 0$) is the Fourier transform of $2a / ( a^2 + \omega^2 )$ within a prefactor.  Following Refs.~\citenum{Duc19,Duc20}, the log scale is used so as to allow a regular sampling of the error oscillations at energies between $E_{min}$ and $E_{max}$. The problem can be then solved in a traditional least square approach using a uniform sampling in $u$. Such a formulation conserves an excellent accuracy, as demonstrated in Table~\ref{Table:wtgrids}, \red{with a number of grid points comparable to that commonly adopted with the more elaborated  minimax approach.} \cite{Liu16, Kaltak14,Kaltak14b}
\red{Similarly to the minimax formulation, the grids points $\tau_p$ have been pre-tabulated with the $E_{max}/E_{min}$ ratio as a single parameter, so as to minimize the computational effort of the setup. On the other hand, the $\omega^p_k$ coefficients can be conveniently recalculated on the fly as the result of a simple linear least square equation.}
In association with $n\omega$=12 imaginary   frequencies, $n\tau$=18  times are selected to reach sub-meV accuracy on the quasiparticle energies. 
We provide in Table~\ref{Table:wtgrids} a typical test of accuracy, selecting the def2-TZVP $G_0W_0$@PBE0 HOMO and LUMO energies of acridine, the first element of the molecular set of Ref.~\citenum{Knight_2016}  studied in full details in the next section. 
In this Table, RI-RS without Laplace transform only differs from the standard Coulomb-fitting (RI-V) by the construction of the $\mathcal{F}_{\mu}$ fitting coefficients. We observe in particular that the dependence on the n$\omega$ grid falls well below the meV for n$\omega \ge$ 10. The real-space approach with  Laplace-transform (RI-RS+LT) depends further on the imaginary-time grid. However, for a given n$\omega$ imaginary-frequency grid, a time-grid with n$\tau =1.5 \times n\omega$ introduces negligible errors, comforting overall our choice of n$\omega=12$ and n$\tau$=18 running parameters.

 


\section{Results}


\subsection{Validation and accuracy}

We benchmark the accuracy of the present scheme  using the recent set of 24 intermediate size molecules with acceptor character proposed in Ref.~\citenum{Knight_2016}. Our calculations are performed at the def2-TZVP $G_0W_0$@PBE0 level associated with the corresponding def2-TZVP-RI auxiliary basis.\cite{Weigend98}  Our goal here is not to carry calculations in the complete-basis set limit, but rather to assess the accuracy of the present space-time approach, as compared to the standard Coulomb-fitting (RI-V) scheme, using a reasonable basis set. The real-space $\lbrace {\bf r}_k \rbrace$ sets were thus  optimized for the def2-TZVP and def2-TZVP-RI basis sets association, following the scheme described above and summarized in Eqn.~\ref{fitting}. As discussed above, the size of the $\lbrace {\bf r}_k \rbrace$ atomic distributions amounts to 136 for H and 336 for second row elements.

\begin{table}
\begin{tabular}{l|rr|rr}
                     &   \multicolumn{2}{c}{HOMO}       &  \multicolumn{2}{c}{LUMO} \\
                     &     RI-V [eV] & RI-RS [eV] &  RI-V [eV] & RI-RS [eV]  \\ 
  \hline
                      anthracene     &     -7.0787   &     -7.0787    &    -0.4233    &    -0.4234     \\
                        acridine     &     -7.5611   &     -7.5611    &    -0.7604    &    -0.7604     \\
                       phenazine     &     -7.9755   &     -7.9754    &    -1.1796    &    -1.1799     \\
                         azulene     &     -7.1271   &     -7.1272    &    -0.5595    &    -0.5597     \\ 
               benzoquinone (BQ)     &     -9.7275   &     -9.7273    &    -1.6004    &    -1.6006     \\
                naphthalenedione     &     -9.3297   &     -9.3296    &    -1.5520    &    -1.5521     \\
                        dichlone     &     -9.4112   &     -9.4111    &    -1.9812    &    -1.9814     \\
                          F4-BQ      &    -10.5855   &    -10.5853    &    -2.3221    &    -2.3220     \\
                         Cl4-BQ      &     -9.7245   &     -9.7246    &    -2.5192    & {\bf -2.5191} \\
                    nitrobenzene     &     -9.7868   &     -9.7867    &    -0.5485    &    -0.5486     \\
        F4-benzenedicarbonitrile     &    -10.2336   &    -10.2334    &    -1.7207    & {\bf -1.7210} \\
            dinitro-benzonitrile     &    -10.7596   &    -10.7595    &    -1.8683    &    -1.8684     \\
              nitro-benzonitrile     &    -10.2038   &    -10.2038    &    -1.3992    &    -1.3993     \\
                    benzonitrile     &     -9.5108   &     -9.5107    &     0.1831    &     0.1828     \\
                   fumaronitrile     &    -10.9712   &    -10.9710    &    -1.0740    &    -1.0740     \\
                           mDCNB     &    -10.0173   & {\bf -10.0175} &    -0.7227    &    -0.7228     \\
                            TCNE     &    -11.4676   &    -11.4674    &    -3.2543    &    -3.2544     \\
                            TCNQ     &     -9.1373   &     -9.1373    &    -3.5795    &    -3.5795     \\
                maleic-anhydride     &    -10.7783   &    -10.7782    &    -1.0029    &    -1.0031     \\
                     phthalimide     &     -9.6757   &     -9.6755    &    -0.6380    &    -0.6381     \\
              phthalic-anhydride     &    -10.1111   &    -10.1111    &    -0.9060    &    -0.9061     \\
         Cl4-isobenzofuranedione     &     -9.5809   &     -9.5807    &    -1.7124    &    -1.7127     \\
                            NDCA     &     -8.7061   & {\bf -8.7058}  &    -1.3302    &    -1.3304     \\
                          BODIPY     &     -7.8008   &     -7.8008    &    -1.6315    &    -1.6316     \\
  \hline   
 Max. err.  &   \multicolumn{2}{c|}{\textbf{+0.32} meV / \textbf{-0.23} meV}    &  \multicolumn{2}{c}{\textbf{+0.07} meV / \textbf{-0.33} meV}  \\
 MAE        &    \multicolumn{2}{c|}{0.12  meV}         &   \multicolumn{2}{c}{0.14  meV}     \\
 MSE        &    \multicolumn{2}{c|}{0.08  meV}        &     \multicolumn{2}{c}{0.12  meV}  \\
\end{tabular}
  \caption{ HOMO and LUMO energies at the def2-TZVP $G_0W_0$@PBE0 level for the molecular set of Ref.~\citenum{Knight_2016}. Molecules are arranged by chemical families in the order of Ref.~\citenum{Knight_2016}. The standard Coulomb-fitting scheme (RI-V) performed with the def2-TZVP-RI auxiliary basis set \cite{Weigend98} is compared to the present real-space Laplace-transform (RI-RS) calculations. Negative and positive maximum errors, the mean absolute (MAE) and mean signed (MSE) errors are indicated in meV. Values leading to the largest error are in bold. } 
\label{Table:maromset}
\end{table}

We provide in Table~\ref{Table:maromset} the def2-TZVP $G_0W_0$@PBE0 highest occupied (HOMO) and lowest unoccupied (LUMO) molecular orbitals energies  calculated using the standard Coulomb-fitting (RI-V) scheme   and the separable real-space (RI-RS) approach in conjunction with the Laplace transform scheme. Both calculations are performed with the ``robust" analytic continuation (AC) scheme.\cite{Duc20} Direct contour-deformation calculations without any analytic continuation, namely calculating directly the needed residues of $W$ along the real-axis within the standard quartic scaling RI-V formalism, shows that the errors introduced by the AC are well below the meV for the HOMO and LUMO energy levels of the molecules contained in this set. The analysis of the results evidences that for such $\lbrace {\bf r}_K \rbrace$ distributions, the error on the quasiparticle energies remains below the meV. Such an accuracy may be tuned by increasing/decreasing the size of the  $\lbrace {\bf r}_k \rbrace$ distributions, but the present accuracy-to-size trade-off is already excellent in practice. 

\begin{figure}[t]
	\includegraphics[width=0.6\linewidth]{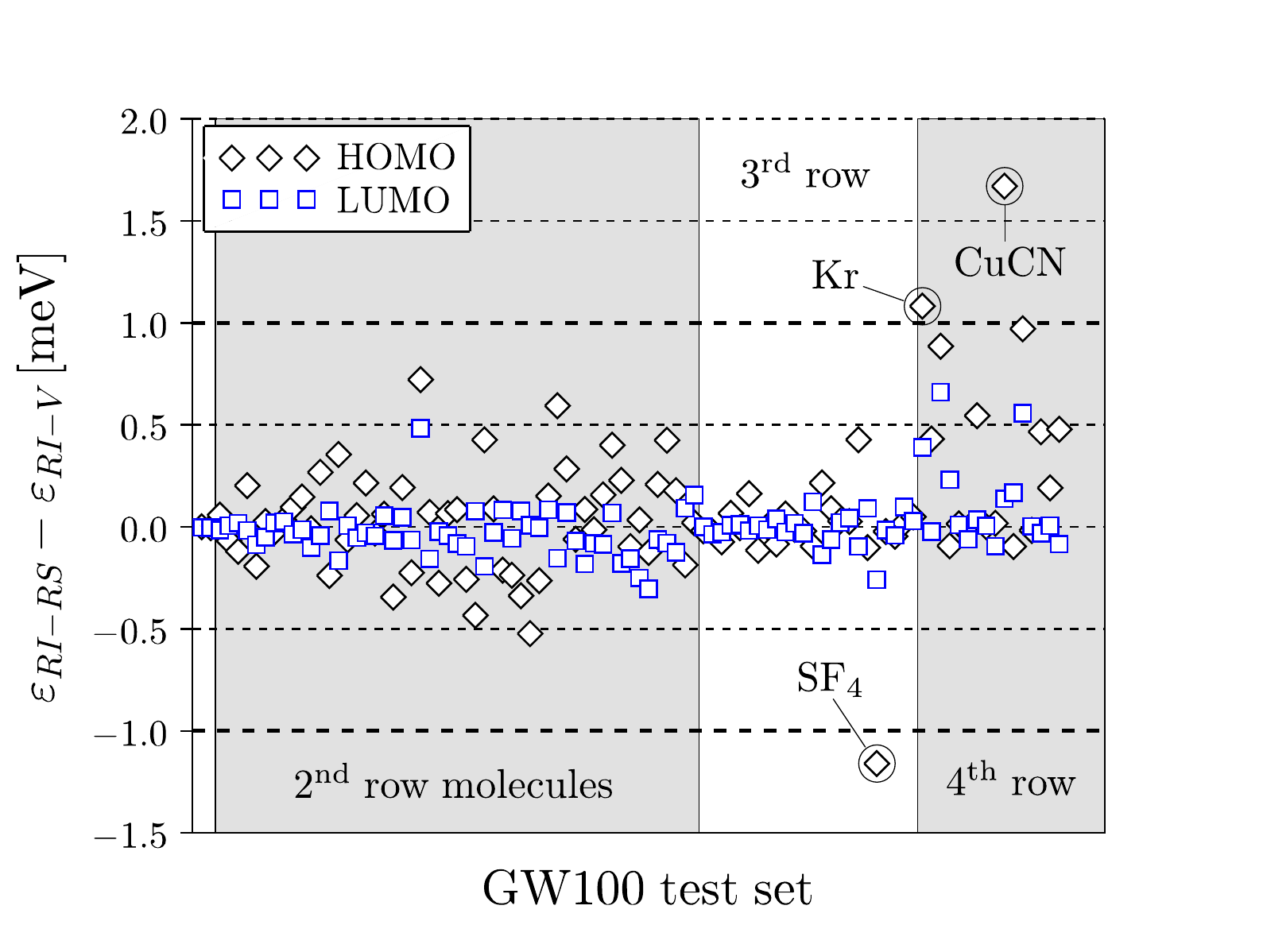}
	\caption{
	HOMO/LUMO def2-TZVP $\mathrm{G_0W_0@PBE}$ quasi-particle energy discrepancy analysis over the $GW$100 test, excluding the 5 systems containing 5-th period elements (see text). The error is that of the present real-space Laplace-transform (RI-RS) approach with respect to the standard Coulomb-fitting (RI-V) approach performed with the corresponding def2-TZVP-RI auxiliary basis set. \cite{Weigend98}  The molecules of the set are sorted according to the maximum period (row) involved within the periodic table. 
	\label{fig_GW100}}
\end{figure}

We further perform benchmark def2-TZVP $G_0W_0$@PBE0 calculations on the $GW$100 test set  \cite{Setten_2015,Krause15,Caruso_2016,Maggio17,Govoni18,Duc20,Gao20} that contains elements from the third and fourth periods of the periodic table,   including transition metal complexes. We exclude the 5 systems containing 5-th period elements for which the def2-TZVP basis set requires the use of an effective core potential, namely Xe, Rb$_2$, I$_2$, vinyl iodide (C$_2$H$_3$I) and  aluminum iodide (AlI$_3$).  Similarly to the previous test set, the error induced by the real-space RI-RS with Laplace transform technique, as compared to a standard RI-V calculations, is   below the meV for most molecules as reported in Fig.~\ref{fig_GW100}. Three systems (Kr, CuCN, SF$_4$) show an error on the HOMO slightly larger than 1 meV (in absolute value), while all errors on the LUMO value are below the meV. As emphasized above, optimizing further the distribution of $\lbrace  {\bf r}_k \rbrace$  points may bring all errors below the meV, but the purpose of the present study is to show that the present scheme, as it stands, already brings very consistently  the error at the meV level. Overall,  the mean absolute (MAE) errors  amount to 0.21~meV and 0.09~meV   for the  HOMO/LUMO, respectively. All data can be found in the Supporting Information. The present data confirm, in the specific case of $GW$ calculations, previous studies reporting on the excellent accuracy-to-cost ratio associated with grid-based techniques for the evaluation of exact exchange or explicit  correlation energies in the context of all-electron atomic-orbital basis sets calculations. \cite{Neese09,Neese11,Klopper13,Parrish12,Todd12,Todd2015,Rebolini16,Duc19,Matthews20} 

\subsection{Scaling analysis}

We finally address the issue of scaling with respect to the system size through the example of  finite-size hexagonal boron-nitride (\textit{h}-BN) ``flakes" containing a central point-defect. The study of the optical emission mediated by defects in  \textit{h}-BN  is an important technological research area, with the prospect of having at hand stable, room-temperature, polarized and ultrabright single-photon sources, together with a scientific challenge when it comes to identify  the defects and mechanisms responsible for such sharp emission lines in the visible range. \cite{Tran_2015,Tran_2016,Martinez_2016,Bourrellier_2016,Jungwirth_2017}

We select as a test case  the $C_BV_N$ (nitrogen vacancy plus carbon substitution to neighbouring boron) defect that has been recently identified as a possible candidate for emission at about 2 eV.\cite{Feng_2017} Our goal here is not to confirm the likeliness of such a defect, but rather to start exploring whether many-body calculations with a typical defect can be performed using finite-size clusters, rather than the traditional supercell approach using periodic boundary conditions (PBC). 
Indeed, the use of PBC complicates the calculations of charged excitations due to the electrostatic interaction between cells, and the Coulomb potential must be truncated to avoid spurious contributions even in the limit of large supercells.    
As such, the modeling of the opto-electronic properties of defects in \textit{h}-BN at the many-body $GW$ and Bethe-Salpeter level  remains scarce due in particular to the cost of performing the required large-scale $GW$ calculations. \cite{Attaccalite_2011,Feng_2017}

\begin{figure}[t]
	\includegraphics[width=0.6\linewidth]{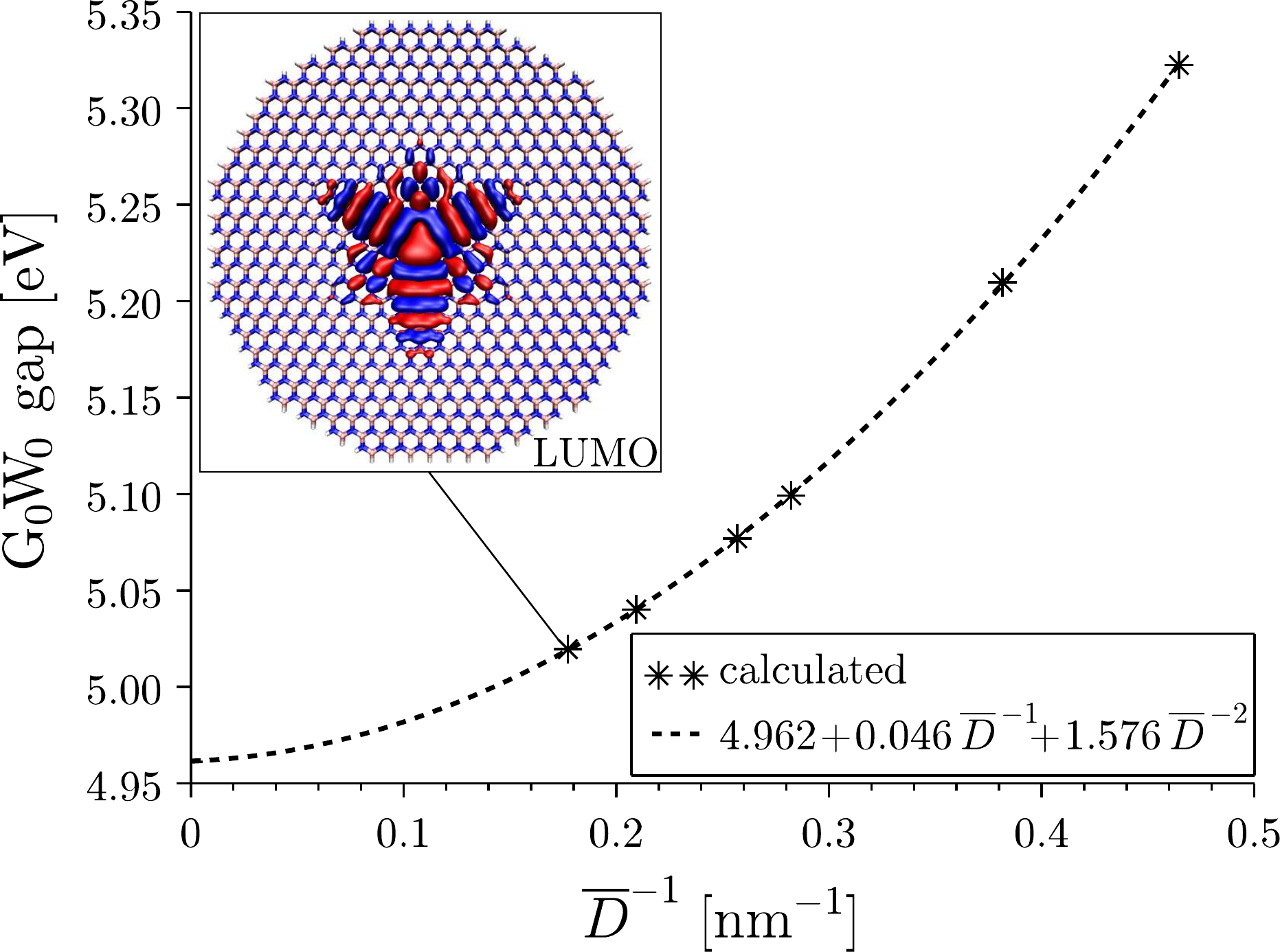}
	\caption{Plot of the 6-311G* $G_0W_0$@PBE0 HOMO-LUMO gap as a function of the inverse flake average diameter. The Inset represents the Kohn-Sham LUMO associated with our largest flake (1019 atoms).
	\label{fig:BNgap}}
\end{figure}

The edge of the \textit{h}-BN flakes are passivated by hydrogen atoms to avoid dangling bonds and the HOMO-LUMO gap is clearly controlled by very localized defect states yielding energy levels within the gap of pristine \textit{h}-BN (see Inset Fig.~\ref{fig:BNgap} for the LUMO). The size of the studied flakes correspond to average radii ranging from 21.5 to 56.4~\AA, containing from 137 to 941 C, B or N atoms, that is  from 167 to 1019 atoms including passivating H atoms.  The average diameter is defined as
$\overline{D} = 2\sqrt{N_{at}S_{at}/ \pi},$ where $N_{at}$ is the number of B or N atoms, and $S_{at}=3\sqrt{3}d_{BN}^2/4$ is the effective surface per B or N atom in the hexagonal lattice, with $d_{BN}$ the BN bond length. Structural relaxation at the PBE0 6-311G* level  indicates that the ground-state for these systems is not spin-polarized.

Before discussing scaling properties, we briefly comment on the evolution of the HOMO-LUMO   energy gap obtained at the 6-311G* $G_0W_0$@PBE0 level \cite{Pople80} as a function of system size (see  Fig.~\ref{fig:BNgap}). \red{ Our goal here is not to obtain   converged values with respect to basis completeness, but to study the evolution of the gap with system size using a minimal triple-zeta plus polarization basis, keeping in mind that the 6-311G* HOMO-LUMO gaps for our defected flakes are typically 70 meV larger than that obtained with the larger def2-TZVP basis set.}  
The decrease of the gap with increasing diameter can be attributed to polarization effects, namely the fact that upon calculating the ionization potential or electronic affinity, as measured by a photo-emission experiment, the added charge localized on the defect generates a long-range Coulomb field that polarizes the surrounding atoms. Such a polarization, properly described within the $GW$ formalism,  stabilizes the added hole or electron,   closing the gap. In the case of finite size systems, this polarization is incomplete as compared to an infinite sheet, leading to a HOMO-LUMO gap that is too large. Performing a fit  of the $GW$ gap up to second order in (1/$\overline{D}$), the linear contribution is found to be negligible, leading to a simple quadratic dependence.   Such a quadratic behaviour stems from the reaction field generated by the 2D density of   dipoles induced by a charge added or removed on/from the LUMO/HOMO levels. \cite{2D_polarization}  The extrapolated gap at infinite radius amounts to 4.96 eV, still $\simeq$60 meV away from the gap of the largest flake studied.  As expected, the $G_0W_0$ gap is much larger than the PBE0 Kohn-Sham gap of 3.07~eV obtained for the largest system.   Fitting the data associated with the 4 smallest flakes, the extrapolated value remains within 20 meV of the extrapolated value with the fit containing all points. This indicates the stability of the extrapolation scheme  and suggests that an accurate  asymptotic value may be obtained with calculations performed on systems containing a rather limited number of atoms.  
While the isolated defect limit $GW$ quasiparticle gap needs extrapolating to infinite sizes, preliminary results indicate that optical excitations, which are neutral excitations of the system, converge much faster as a function of system size. 

\begin{figure}[t]
	\includegraphics[width=0.6\linewidth]{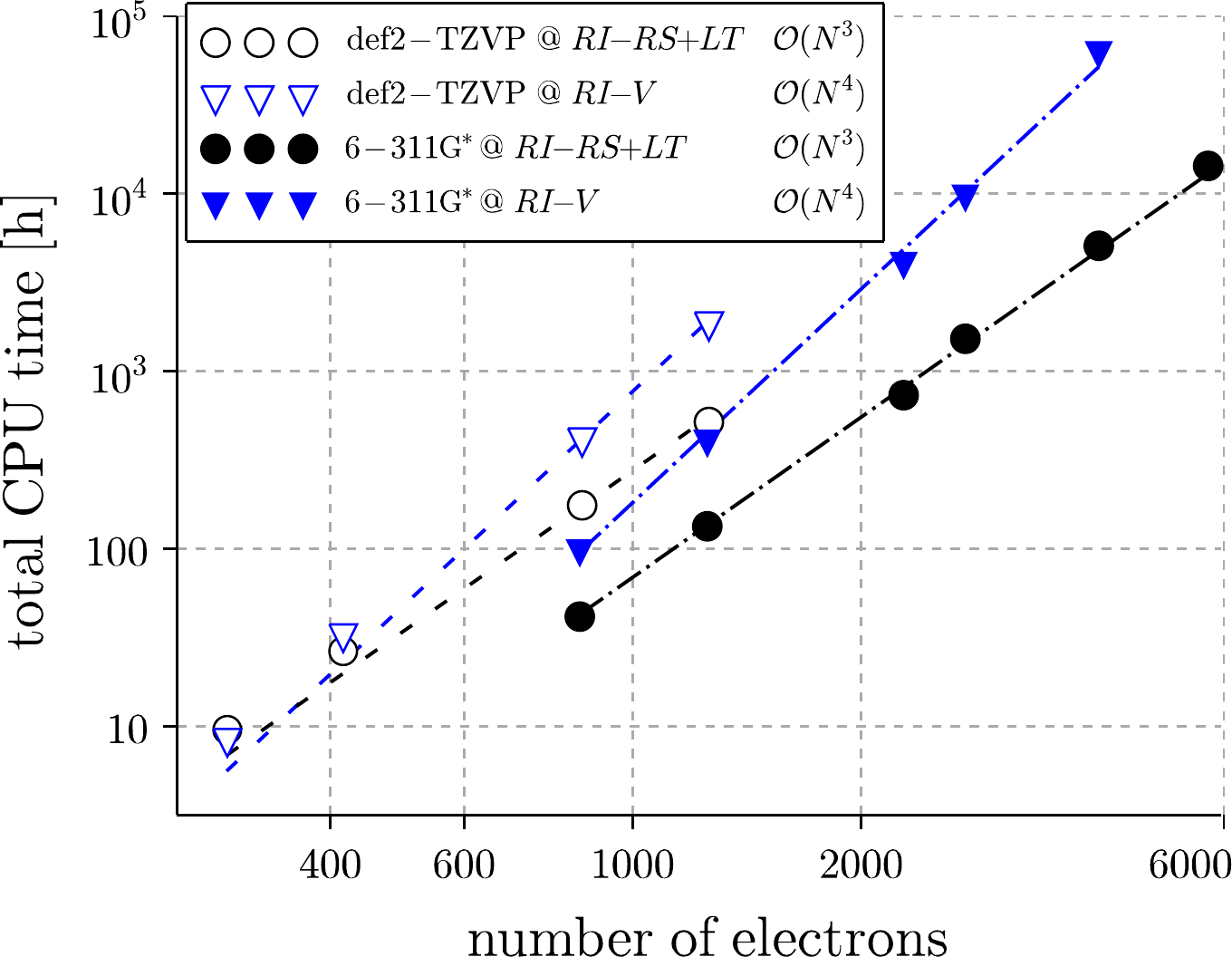}
	\caption{Scaling properties (log scale) for the standard Coulomb-fitting (RI-V) $G_0W_0$ calculations compared with the present real-space Laplace-transform (RI-RS+LT) scheme. Calculations have been performed with the def2-TZVP and 6-311G* basis sets. For the standard RI-V scheme, the auxiliary  def2-TZVP-RI \cite{Weigend98} and universal Coulomb fitting \cite{Weigend2006}   basis sets, respectively, were used. Calculations were performed on a set of hexagonal boron-nitride flakes with up to 6000 electrons.
	Calculations have been performed on AMD Rome Epyc nodes with 128 cores/node and 1.85  Gb/core. \red{ The black (RI-RS+LT) and blue (RI-V) dot-dashed lines are cubic and quartic fits, respectively. An unconstrained fit yields a scaling exponent of 3.07 for the (RI-RS+LT) scheme. }
	\label{fig:scaling}}
\end{figure}

We now plot (log scale) in Fig.~\ref{fig:scaling} the total CPU time (namely the sum of all cores CPU time, over the complete run) associated with the $G_0W_0$ calculations reported in Fig.~\ref{fig:BNgap}. We compare in particular the standard RI-V (Coulomb fitting) calculations (filled blue triangles) performed with the universal Coulomb fitting auxiliary basis of Ref.~\citenum{Weigend2006} and the present real-space Laplace-transform approach (filled black circles)  with a  real-space $\lbrace {\bf r}_k \rbrace$ distribution optimized as described above for this specific auxiliary basis set. 
\red{Timings are reported in Tables S2 and S3 of the Supporting Information where the specific contributions from the RI-RS set-up, susceptibility calculations, Dyson-equation inversion and self-energy calculations are further provided. }

These calculations confirm that the present space-time scheme offers a cubic scaling with system size (see black dot-dashed fit) with an (extrapolated) crossover with the standard quartic scaling RI-V scheme taking place for about 350 electrons.   
We further explore limit of small system sizes with a larger def2-TZVP basis associated with its def2-TZVP-RI auxiliary basis. We can see that despite the larger basis sets, the overhead of the 128 core distribution and pre-computation phases still hinders slightly the fit by perfect cubic/quartic lines in the small system size limit. 
Nonetheless, these later calculations confirms a crossover that consistently takes place at about 350 electrons ($\simeq$50-60 B/N atoms).
A similar crossover was observed in the case of cc-pVTZ RPA calculations where we used a similar real-space Laplace-transform approach  to build the independent-electron susceptibility (see Ref.~\citenum{Duc19}). 


All calculations have been performed on a supercomputer built of 128 cores 2,6 GHz AMD Rome nodes with \red{1.85~Gb}/core memory. We only used fully filled nodes in order to maintain consistency between the timings presented here, meaning that the smallest system calculations have been distributed on a minimum of 128 cores. \red{Under this constraint, we selected CPU grid sizes that roughly match the minimum memory requirement for each calculation, as detailed in Table S2 of the Supporting Information. Our real-space Laplace-transform (RI-RS+LT) calculations require much less cores than the standard RI-V scheme, partly due to the corresponding $\mathcal{O}(N^2)$ memory footprint, with the $\chi_0({\bf r}_k,{\bf r}_k'; i\omega)$ being the largest objects stored in memory. On the other hand, the memory requirement of our standard RI-V implementation grows as $\mathcal{O}(N^3)$, dominated in this case by the storage of (occupied)$\times$(virtual) co-density auxiliary fits $\mathcal{F}_{\mu}^V(\phi_i \phi_a)$. Let us emphasize that within the RI-RS+LT approach, each 3-center integral is computed and immediately discarded during the RI setup.} 





\section{Conclusions}

We have presented an all-electron space-time $GW$   formalism relying on  a separable resolution-of-the-identity (RI)  formalism, offering cubic-scaling $GW$ calculations  that do not exploit any sparsity nor localization considerations. This allows a crossover with the quartic scaling Coulomb-fitting RI-V $GW$ calculations for systems containing a very few hundred electrons, independently of  the dimensionality of the studied system. As compared to the  interpolative separable density fitting (ISDF) scheme, the present approach preserves  the use of  standard auxiliary Gaussian basis sets that are taken as an input, and not constructed by the ISDF algorithm. The needed distribution of $\lbrace {\bf r}_k \rbrace$ points 
are optimized to recover at the meV level the results of a standard Coulomb-fitting (RI-V) $GW$ calculation. \red{ Precalculated grids to be associated with a larger collection of standard basis sets, beyond the 6-311G* and def2-TZVP sets adopted in this study, comes now as a prerequisite for a broader use of the present scheme. } 
Scaling with system size could be further reduced possibly by exploiting stochastic techniques \cite{Neu13} or the decay properties of the space-time Green's functions \cite{Schindl00} at long-range in the case of very large systems.  The performances as they stand today  clearly illustrate however the interest of real-space quadrature, as developed by the quantum chemistry community  for all-electron atomic-orbital basis sets calculations, and the   progress performed by the $GW$ community to install this family of many-body perturbation techniques as a valuable tool for  moderately correlated systems, with an excellent trade-off between accuracy and CPU time, allowing to tackle finite size or periodic systems, metallic or semiconducting, containing up to several hundred atoms.  

 \begin{acknowledgement}
 The authors are indebted to Pascal Pochet for discussions concerning defects stability in periodic and finite size systems and to Gabriele D'Avino for explaining the quadratic behaviour of the polarization energy with inverse diameter. Calculations have been performed thanks to an allocation on the French CEA-TGCC Joliot-Curie supercomputer  comprizing
 2292  AMD Rome Epyc nodes with 128 cores/node and 2 Go/core. This work received support from the   French \textit{Agence Nationale de la Recherche} (ANR) under contract ANR-20-CE29-0005.
  \end{acknowledgement}
\begin{suppinfo}

We provide in the Supporting Information the details of the $GW$100 test set data (Table S1) and additional information about the systems size, timings and number of cores used for the 6-311G* $G_0W_0$@PBE0 calculations on the hexagonal boron-nitride flakes (Tables S2-S3 and Fig.~S1).

\end{suppinfo}

\bibliography{biblio}

\begin{tocentry}
 \includegraphics[width=\linewidth]{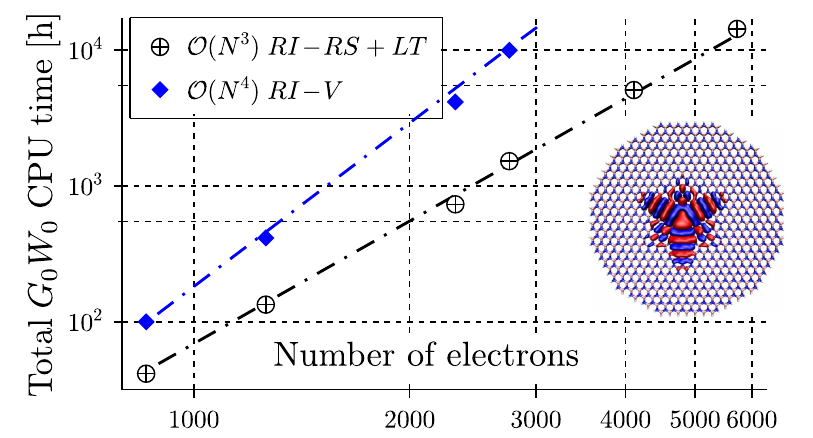}   
\vskip .2cm  
Scaling with system size for 6-311G* $G_0W_0$ calculations performed with the cubic scaling real-space Laplace transform scheme (RI-RS+LT) as compared to the quartic scaling traditional Coulomb fitting (RI-V)  approach for a family of defected boron-nitride flakes. 
\end{tocentry}

\end{document}





\section{Details of the def2-TZVP $G_0W_0$ calculations on the $GW$100 test set. }

\begin{spacing}{1.0}
\begin{longtable}{l|rr|rr}
\caption{ HOMO and LUMO energies at the def2-TZVP $G_0W_0$@PBE0 level for the $GW$100 molecular test set. \cite{Setten_2015}    The standard Coulomb-fitting scheme (RI-V) performed with the def2-TZVP-RI auxiliary basis set \cite{Weigend98} is compared to the present real-space Laplace-transform (RI-RS) calculations. Negative and positive maximum errors, the mean absolute (MAE) and mean signed (MSE) errors are indicated in meV. Values leading to the largest error are in bold. } \\ \label{Table:GW100}

  &   \multicolumn{2}{c}{HOMO}       &  \multicolumn{2}{c}{LUMO} \\
  &     RI-V [eV] & RI-RS [eV] &  RI-V [eV] & RI-RS [eV]  \\ 
  \hline
\endfirsthead  

\multicolumn{5}{c}%
{{\bfseries \tablename\ \thetable{} -- GW100 $G_0W_0$@PBE0 data, continued from previous page}} \\ \multicolumn{5}{c}{} \\
  &   \multicolumn{2}{c}{HOMO}       &  \multicolumn{2}{c}{LUMO} \\
  &     RI-V [eV] & RI-RS [eV] &  RI-V [eV] & RI-RS [eV]  \\ 
  \hline
\endhead

\hline \multicolumn{5}{r}{{Continued on next page}} \\
\endfoot

\endlastfoot

                    Acetaldehyde &    -9.8658 &    -9.8654 &     1.7301 &     1.7299 \\ 
                       Acetylene &   -11.1190 &   -11.1189 &     3.4524 &     3.4524 \\ 
                         Adenine &    -8.0278 &    -8.0277 &     1.0619 &     1.0617 \\ 
             Aluminumtrifluoride &   -14.7007 &   -14.7006 &     0.6740 &     0.6739 \\ 
                          Amonia &   -10.4839 &   -10.4837 &     3.0423 &     3.0423 \\ 
                         Aniline &    -7.6936 &    -7.6934 &     1.6764 &     1.6762 \\ 
                           Argon &   -15.1917 &   -15.1917 &    14.7162 &    14.7162 \\ 
                    Arsenicdimer &    -9.4097 &    -9.4093 &    -0.3135 &    -0.3136 \\ 
                          Arsine &   -10.1110 &   -10.1111 &     3.0944 &     3.0946 \\ 
                         Benzene &    -9.0143 &    -9.0140 &     1.5996 &     1.5997 \\ 
               Berylliummonoxide &   -12.9594 &   -12.9595 &   -11.6442 &   -11.6442 \\ 
                          Borane &   -13.0064 &   -13.0064 &     0.4548 &     0.4547 \\ 
                    Boronnitride &   -11.3781 &   -11.3782 &    -3.5945 &    -3.5945 \\ 
                         Bromine &   -10.1504 &   -10.1495 &    -0.8253 &    {\bf -0.8247} \\ 
                         Buthane &   -11.5753 &   -11.5754 &     2.9304 &     2.9302 \\ 
                   Carbondioxide &   -13.4373 &   -13.4371 &     2.9403 &     2.9403 \\ 
                 Carbondisulfide &    -9.7561 &    -9.7562 &     0.2424 &     0.2424 \\ 
                  Carbonmonoxide &   -13.9578 &   -13.9576 &     1.0780 &     1.0779 \\ 
               Carbonoxyselenide &   -10.1634 &   -10.1633 &     1.3249 &     1.3248 \\ 
                Carbonoxysulfide &   -10.9454 &   -10.9453 &     1.7287 &     1.7287 \\ 
              Carbontetrabromide &    -9.9729 &    -9.9719 &    -0.4853 &    -0.4847 \\ 
             Carbontetrachloride &   -11.1414 &   -11.1409 &     0.7781 &     0.7780 \\ 
             Carbontetrafluoride &   -15.7426 &   -15.7419 &     4.9648 &     4.9653 \\ 
                        Chlorine &   -11.1315 &   -11.1316 &    -0.1991 &    -0.1991 \\ 
                   Coppercyanide &   -10.0069 &  {\bf -10.0052} &    -1.0328 &    -1.0327 \\ 
                     Copperdimer &    -7.3478 &    -7.3479 &    -0.3626 &    -0.3624 \\ 
               Cyclooctatetraene &    -8.0888 &    -8.0886 &     0.5793 &     0.5791 \\ 
                 Cyclopentadiene &    -8.3697 &    -8.3699 &     1.5707 &     1.5707 \\ 
                    Cyclopropane &   -10.6135 &   -10.6138 &     3.5310 &     3.5311 \\ 
                        Cytosine &    -8.4546 &    -8.4546 &     0.8030 &     0.8029 \\ 
                       Diborane6 &   -11.9952 &   -11.9951 &     1.2461 &     1.2460 \\ 
                     Dipotassium &    -3.9849 &    -3.9849 &    -0.4377 &    -0.4377 \\ 
                        Disilane &   -10.3480 &   -10.3480 &     2.2701 &     2.2702 \\ 
                          Ethane &   -12.4466 &   -12.4468 &     3.1998 &     3.1998 \\ 
                         Ethanol &   -10.3587 &   -10.3592 &     2.9515 &     2.9515 \\ 
                    Ethoxyethane &    -9.5363 &    -9.5364 &     3.0017 &    {\bf 3.0014} \\ 
                     Ethybenzene &    -8.5805 &    -8.5805 &     1.5685 &     1.5687 \\ 
                        Ethylene &   -10.3456 &   -10.3455 &     2.5831 &     2.5831 \\ 
                        Fluorine &   -15.3320 &   -15.3322 &     0.1373 &     0.1372 \\ 
                    Fluoroborane &   -10.7671 &   -10.7671 &     1.5093 &     1.5092 \\ 
                    Formaldehyde &   -10.5488 &   -10.5487 &     1.5744 &     1.5745 \\ 
                      Formic Acid &   -11.0648 &   -11.0647 &     2.5916 &     2.5915 \\ 
             Galliummonochloride &    -9.5438 &    -9.5438 &     0.3533 &     0.3532 \\ 
                         Germane &   -12.1362 &   -12.1362 &     3.2973 &     3.2973 \\ 
                         Guanine &    -7.7447 &    -7.7449 &     1.3749 &     1.3750 \\ 
                          Helium &   -23.7926 &   -23.7926 &    22.2760 &    22.2760 \\ 
               Hexafluorobenzene &    -9.6562 &    -9.6563 &     0.9102 &     0.9101 \\ 
                       Hydrazene &    -9.4265 &    -9.4264 &     2.6692 &     2.6691 \\ 
                        Hydrogen &   -15.9491 &   -15.9491 &     4.4635 &     4.4635 \\ 
                   Hydrogenazide &   -10.4807 &   -10.4811 &     1.8777 &     1.8776 \\ 
                Hydrogenchloride &   -12.2631 &   -12.2632 &     2.8900 &     2.8900 \\ 
                 Hydrogencyanide &   -13.3566 &   -13.3569 &     3.1549 &     3.1550 \\ 
                Hydrogenfluoride &   -15.5491 &   -15.5491 &     3.2673 &     3.2673 \\ 
                Hydrogenperoxide &   -11.2224 &   -11.2222 &     3.1763 &     3.1764 \\ 
                 Hydrogensulfide &   -10.0084 &   -10.0084 &     3.2376 &     3.2376 \\ 
                         Krypton &   -13.4876 &   -13.4865 &    10.4118 &    10.4122 \\ 
                    Lithiumdimer &    -5.0389 &    -5.0389 &    -0.1772 &    -0.1771 \\ 
                 Lithiumfluoride &   -10.4604 &   -10.4603 &     0.1233 &     0.1232 \\ 
                  Lithiumhydride &    -7.3073 &    -7.3072 &     0.1841 &     0.1841 \\ 
               Magnesiumchloride &   -11.2469 &   -11.2469 &    -0.1141 &    -0.1142 \\ 
               Magnesiumfluoride &   -12.9920 &   -12.9921 &     0.0598 &     0.0599 \\ 
               Magnesiummonoxide &    -7.3227 &    -7.3226 &    -1.5708 &    -1.5708 \\ 
                         Methane &   -14.0449 &   -14.0452 &     3.5594 &     3.5594 \\ 
                        Methanol &   -10.7266 &   -10.7268 &     3.1640 &     3.1639 \\ 
                            Neon &   -20.7658 &   -20.7657 &    20.8885 &    20.8885 \\ 
                        Nitrogen &   -15.2455 &   -15.2455 &     2.9042 &     2.9041 \\ 
                            Ozon &   -12.3624 &   -12.3620 &    -1.8652 &    -1.8654 \\ 
                     Pentasilane &    -8.9797 &    -8.9796 &     0.7289 &     0.7289 \\ 
                          Phenol &    -8.3968 &    -8.3966 &     1.4663 &     1.4663 \\ 
                     Phenol (v2) &    -8.2916 &    -8.2912 &     1.5195 &     1.5196 \\ 
                       Phosphine &   -10.2794 &   -10.2793 &     3.1459 &     3.1460 \\ 
                 Phosphorusdimer &   -10.1723 &   -10.1723 &    -0.2715 &    -0.2715 \\ 
           Phosphorusmononitride &   -11.4976 &   -11.4975 &     0.2857 &     0.2857 \\ 
                Potassiumbromide &    -7.6723 &    -7.6718 &    -0.2585 &    -0.2584 \\ 
                Potassiumhydride &    -5.4022 &    -5.4022 &     0.0491 &     0.0492 \\ 
                         Propane &   -11.8751 &   -11.8750 &     3.0092 &     3.0093 \\ 
                        Pyridine &    -9.3883 &    -9.3877 &     1.0332 &     1.0330 \\ 
                          Silane &   -12.4656 &   -12.4657 &     3.2217 &     3.2217 \\ 
                  Sodiumchloride &    -8.5723 &    -8.5725 &    -0.4087 &    -0.4087 \\ 
                     Sodiumdimer &    -4.9076 &    -4.9076 &    -0.3663 &    -0.3663 \\ 
                   Sodiumhexamer &    -4.2910 &    -4.2910 &    -0.6794 &    -0.6794 \\ 
                  Sodiumtetramer &    -4.1678 &    -4.1677 &    -0.7052 &    -0.7052 \\ 
                   Sulferdioxide &   -12.0048 &   -12.0045 &    -0.4411 &    -0.4412 \\ 
             Sulfertetrafluoride &   -12.2741 &  { \bf -12.2753 } &     0.9815 &     0.9813 \\ 
                     Tetracarbon &   -11.0137 &   -11.0139 &    -2.5154 &    -2.5154 \\ 
                         Thymine &    -8.8343 &    -8.8341 &     0.5947 &     0.5946 \\ 
                Titaniumfluoride &   -14.6675 &   -14.6670 &    -0.2178 &    -0.2178 \\ 
                         Tuloene &    -8.6321 &    -8.6316 &     1.5334 &     1.5333 \\ 
                          Uracil &    -9.2341 &    -9.2340 &     0.5442 &     0.5440 \\ 
                            Urea &    -9.6647 &    -9.6650 &     2.4063 &     2.4063 \\ 
                    Vynilbromide &    -8.9477 &    -8.9475 &     1.9401 &     1.9401 \\ 
                  Vynilbromidev2 &    -9.4866 &    -9.4861 &     1.8184 &     1.8183 \\ 
                   Vynilchloride &    -9.7883 &    -9.7883 &     2.0109 &     2.0108 \\ 
                   Vynilfluoride &   -10.2372 &   -10.2377 &     2.7246 &     2.7247 \\ 
                           Water &   -12.1646 &   -12.1646 &     3.0779 &     3.0779 \\ 
  \hline
 Max. err.  &   \multicolumn{2}{c|}{\textbf{+1.67} meV / \textbf{-1.16} meV}    &  \multicolumn{2}{c}{\textbf{+0.66} meV / \textbf{-0.30} meV}  \\
 MAE        &    \multicolumn{2}{c|}{0.21  meV}         &   \multicolumn{2}{c}{0.09  meV}     \\
 MSE        &    \multicolumn{2}{c|}{0.08  meV}        &     \multicolumn{2}{c}{0.001  meV}  \\

\end{longtable}
\end{spacing}

\section{Timings for the 6-311G* $G_0W_0$@PBE0 calculations }

We present in Table S2 additional details on the size, timing (walltime) and number of cores used concerning the   defected h-BN systems. Calculations have been performed on 128-cores 2.6 GHz AMD Rome nodes with 1.85 Gb/core. For consistency between different systems, all calculations used fully filled nodes. 
Notice that real-space Laplace-transform (RI-RS+LT) calculations required much less cores than the standard RI-V scheme as the signature of  reduced memory requirements. 

\begin{table}
\caption{ Number of B,C,N and H atoms, number of electrons, number of cores and timings in hh:mm:ss (walltime) for the 6-311G* $G_0W_0$@PBE0 calculations on defected hexagonal boron-nitride flakes. The memory indicated represent the maximum memory available for the run.}   \label{Table:timings}
\begin{tabular}{c|c|c|c|c|c|c|c|c}
\multicolumn{3}{c|}{} & \multicolumn{3}{c|}{ RI-V   } & \multicolumn{3}{c}{ RI-RS+LT   } \\
\hline
  N$_{B/C/N}$ & N$_H$ & N$_{el}$ & N$_{CPU}$ & walltime  & mem.(Tb) & N$_{CPU}$ & walltime & mem.(Tb) \\
  \hline
 137 &  30  &  852 &   384 & 00:15:37 &  0.72 &  128 & 00:19:26 & 0.24 \\ 
 203 &  36  & 1254 &   768 & 00:32:21 &  1.44 &  256 & 00:31:21 & 0.48 \\
 371 &  48  & 2274 &  3072 & 01:21:12 &  5.76 &  640 & 01:08:41 & 1.20 \\
 448 &  54  & 2742 &  6144 & 01:37:25 & 11.52 & 1152 & 01:19:18 & 2.16 \\
 674 &  65  & 4108 & 23040 & 02:46:32 & 43.20 & 2560 & 01:59:11 & 4.80 \\
 941 &  78  & 5724 &   --  &    --    &  --   & 5120 & 02:48:00 & 9.60 \\
\end{tabular}
\end{table}

We further give the total CPU time associated with the main cubic-scaling operations (Fig.~S1 and Table S3), namely the construction of the grid weights $M_{\mu,k}$ (RI setup), the construction of the independent-electron susceptibilities ($\chi_0$), the inversion of the Dyson equation to obtain the screened Coulomb potential ($W$), and finally the evaluation of the self-energy  ($\Sigma^{GW}$). Taking the largest system, for which the 6-311G* space-time $G_0W_0$ calculation represents a total CPU time of about 14336 hours, the sum of these 4 steps amounts to 13984 hours, namely 98$\%$ of the total. 
For this largest flake, the RI set-up represents about 25$\%$ and the $\chi_0$ calculations about 32$\%$ of the total CPU time. \\

\begin{table}
\caption{ Number of atoms,   electrons,   cores and timings in hh:mm (total CPU time) for the principal computational steps of the 6-311G* RI-RS+LT $G_0W_0$@PBE0 calculations.}   \label{Table:timings_details}
\begin{tabular}{c|c|c|c|c|c|c|c}
\multicolumn{4}{c|}{} & \multicolumn{4}{c}{total CPU time} \\
\hline
  N$_{B/C/N}$ & N$_H$ & N$_{el}$ & N$_{CPU}$ & RI setup  & $\chi_0$ & $W$ & $\Sigma_{GW}$ \\
  \hline
 137 &  30  &  852 &   128 & 11:37 & 10:58  & 2:09  & 14:52 \\ 
 203 &  36  & 1254 &   256 & 33:56 & 39:40 & 6:45 & 49:31  \\
 371 &  48  & 2274 &  640 & 171:42  & 202:50  & 39:37  & 301:34 \\
 448 &  54  & 2742 &  1152 & 338:16  & 466:55 & 88:04 & 592:49 \\
 674 &  65  & 4108 & 2560 & 1219:20 & 1555:57 & 241:13  & 1866:17 \\
 941 &  78  & 5724 & 5120 & 3633:19  & 4564:49  & 715:42  & 5070:11  \\
\end{tabular}
\end{table}

\begin{figure}[t]
	\includegraphics[width=0.5\linewidth]{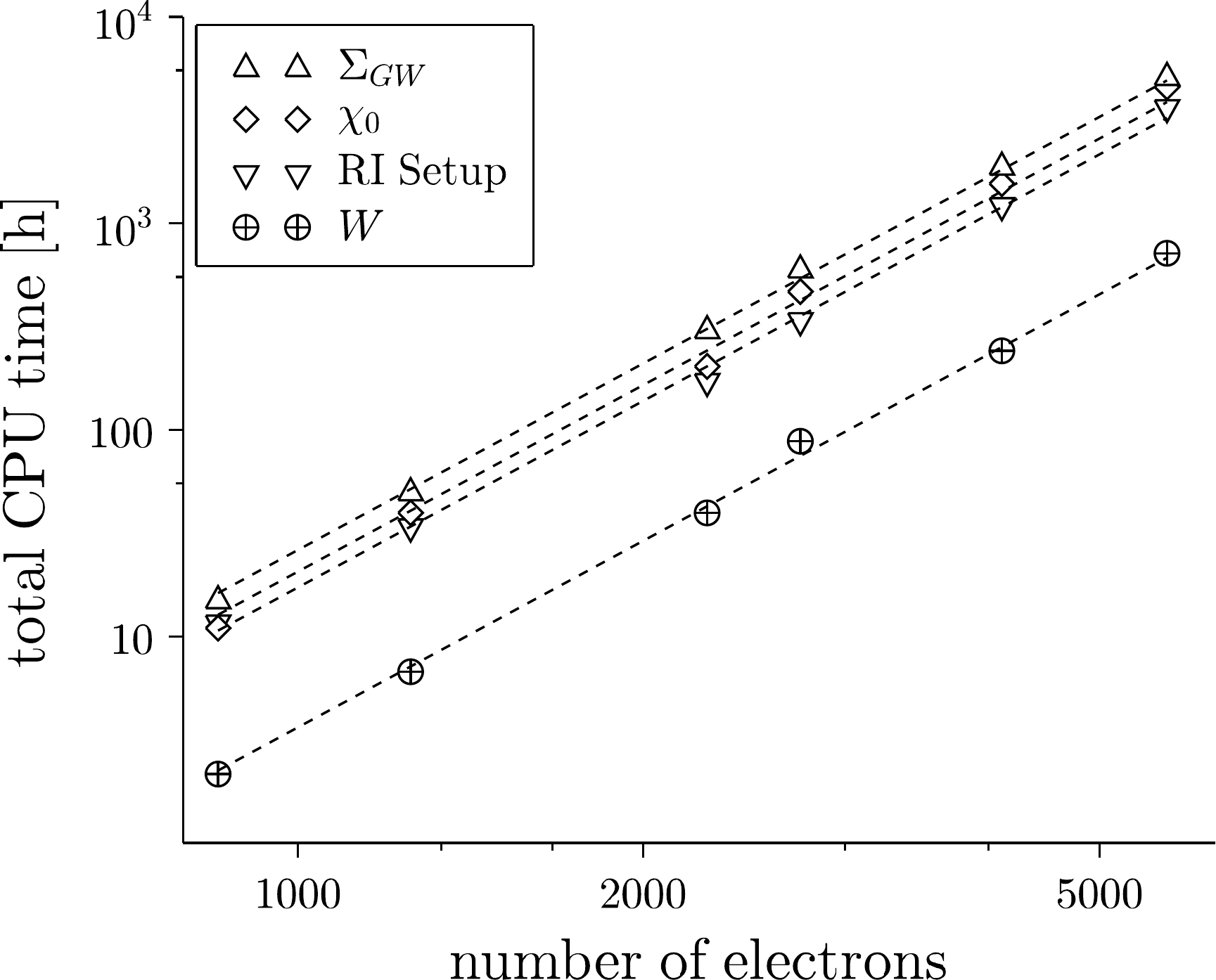}
	\caption{Scaling behavior for the 4 principal steps of our cubic-scaling $G_0W_0$ implementation. Dashed lines are guides for the eyes using an exact cubic behavior (slope equal to 3). 
	\label{fig_timings_details}}
\end{figure}

\vskip 8cm

\bibliography{biblio}